\documentclass[fleqn,usenatbib]{mnras}

\usepackage{newtxtext,newtxmath}

\usepackage[T1]{fontenc}

\DeclareRobustCommand{\VAN}[3]{#2}
\let\VANthebibliography\thebibliography
\def\thebibliography{\DeclareRobustCommand{\VAN}[3]{##3}\VANthebibliography}

\usepackage{graphicx}	
\usepackage{amsmath}	

\usepackage{physics}
\usepackage{multirow}
\usepackage{color,soul,framed,xcolor} 

\title[X-ray variations in $\zeta$ Pup do not require mass flux changes]{Observed epochal variations in X-ray lines from the O Supergiant $\zeta$ Puppis do not require substantial changes in the wind mass flux}

\author[S. J. Gunderson et al.]{Sean J. Gunderson$^{1,2}$\thanks{Contact e-mail: \href{mailto:seang97@mit.edu}{seang97@mit.edu}},
Kenneth G. Gayley$^2$,
David P. Huenemoerder$^1$,
Pragati Pradhan$^3$,
Nathan A. Miller$^4$
\\
$^1$Massachusetts Institute of Technology, Kavli Institute for Astrophysics and Space Research, 77 Massachusetts Ave., Cambridge, MA 02139, USA\\
$^2$Department of Physics and Astronomy, University of Iowa, Iowa City, IA 52242, USA\\
$^3$Embry Riddle Aeronautical University, Department of Physics Prescott Campus, 3700 Willow Creek Road, Prescott, AZ 86301\\
$^4$Department of Physics and Astronomy, University of Wisconsin–Eau Claire, Eau Claire, WI 54701, USA\\
}

\date{Accepted XXX. Received YYY; in original form ZZZ}

\pubyear{2023}

\defcitealias{Cohen10}{C2010}
\defcitealias{Cohen20}{C2020}
\defcitealias{Gunderson22}{G2022}

\begin{document}
\label{firstpage}
\pagerange{\pageref{firstpage}--\pageref{lastpage}}
\maketitle

\begin{abstract}

We fit the high resolution \textit{Chandra} X-ray spectra of the O supergiant $\zeta$ Puppis using the variable boundary condition (VBC) line model to test the stability of its mass-loss rate between two epochs of observation: 2000 March and 2018 July -- 2019 August. At issue is whether the observed variations are induced by global changes in the cool (unshocked) wind itself or are isolated to the local pockets of hot gas (i.e., changes in the frequency and location of the shocks).  Evidence in the literature favored the possibility of a 40 per cent increase in the mass flux of the entire stellar wind, based on X-ray reabsorption from a line-deshadowing-instability-inspired parameterization, whereas our fit parameters are consistent with a constant mass flux with a change in the velocity variations that determine the locations where shocks form. Our results suggest the shocks in the more recent data are formed at somewhat larger radii, mimicking the enhanced blueshifts and increased line fluxes interpreted in the previous analysis as being due to increases in both the X-ray generation and reabsorption from an overall stronger wind.
\end{abstract}

\begin{keywords}
stars: mass-loss -- stars: massive -- X-rays: stars -- stars: winds, outflows -- line: profiles
\end{keywords}



\section{Introduction}

At only 332 pc away \citep{Howarth19}, $\zeta$ Puppis (HD 66811) is one of the brightest O stars in the night sky. This has made it the the canonical O supergiant and one of the most well studied single massive stars across all wave bands. Its well studied nature has also made it an ideal calibration target for some of our most sensitive X-ray instruments (e.g., \textit{XMM-Newton's} RGS and EPIC\footnote{\url{https://xmm-tools.cosmos.esa.int/external/xmm_user_support/documentation/uhb/routinecal.html}}). As a massive star, $\zeta$ Pup's stellar wind is relatively well understood to be driven by radiative forces acting on hypersonically Doppler shifted UV line opacity from metal ions in its atmosphere, As such, it is expected to have a wind that, despite being clumpy \citep{Martinez17}, should on the whole be as steady as its luminosity. This is true even in the case of the star being a rapid rotator, as is the case for $\zeta$ Pup. The source of the high rotation rate, along with its runaway status, is thought to be evidence of a prior companion in $\zeta$ Pup's evolutionary history whose supernova resulted in $\zeta$ Pup's high speed ejection \citep{Woermann01}.

While many of $\zeta$ Pup's stellar properties have been well constrained, a number of studies have also revealed that our picture is perhaps incomplete. Observations in both the optical (\citet{Ram18} using \textit{BRITE}) and X-ray (\citet{Naze18} using \textit{XMM-Newton} and \textit{Swift} and recounting earlier documented variabilities, and \citet{Nichols21} using \textit{Chandra}) regimes have highlighted apparent variabilities in the star. The main periodicity detected in the optical band has a period of 1.78 days (originally detected by \citet{Howarth14}) and been argued to arise from bright spots on $\zeta$ Pup's surface and its rapid rotation \citep{Ram18,Nichols21}. There is also an apparent 2.56-day periodicity \citep{Marcheko98} that \citet{Howarth19} corroborated in the \textit{Hipparcos} data, which interestingly does not show evidence of the aforementioned 1.78-day signal.

Adding another layer of complexity, the periodicities do not appear to be constant. \citet{Naze18} noted that the X-ray periodicity appear to change in phase and period across many \textit{XMM-Newton} observations, at times correlating with the \textit{BRITE} signal and other times not. The strength of the X-ray signal also appears to wax and wane: it was stronger during the 2007-2011 than before or after. The optical 1.78-day signal also appears to be varying as \citet{Howarth19} posited that its amplitude has grown with time to explain why \textit{Hipparcos} does not show it. Older observations reported longer period variations than those today. These include periods of 5.21 days \citep{Balona92} and 5.075 days \citep{Moffat81} using optical data, and 5.1 days \citep{Howarth95} in the UV. How these longer periods fit into the picture is less clear. They were originally proposed to be connected to the rotation of $\zeta$ Pup, but the rotation period is faster \citep[<3 days;][]{Howarth19} than these signal periods. Thus there appears to be multiple layers of periodicity which are themselves varying in frequency, amplitude, and phase relationship between various wavebands.

There is an important distinction to be made between X-ray variability and variability in other wavebands. As a massive star, $\zeta$ Pup generates X-rays not at its static photosphere but farther out in its hypersonic wind through embedded shocks \citep{Lucy80,Feldmeier97} as clumps experience differential acceleration. This X-ray generating mechanism is dependent on the number, size, speed, and many other factors of the wind clumps, so it has an inherent stochasticity. This randomness is associated with the flow time of the wind $t_\mathrm{flow}=R_*/v_\infty$, which is on the order of a couple of hours, a much shorter timescale than those discussed above. An example of a possible short-period signal is the 0.694-day period \citet[reported as 1.44 cycles/day]{Berghoefer96} found in simultaneous X-ray and H$\alpha$ observations. X-ray observations are typically long enough compared with this timescale to average out any stochastic effects (see Table~\ref{tab:ObsLog}).  The noted X-ray periodicities are more coherent signals that persist through the time averaging, so changes in the X-ray signal may correspond to global changes in the time-averaged wind.

In summary, many recent investigations of the relationship between optical and X-ray variations of this star on short time scales have revealed interesting links between the two wave bands. Hence it is timely to ask if there exists longer term changes, either in pockets of X-ray generating shocks, or more globally over the cool wind. Capitalizing on the decades of high resolution X-ray profile information that exists for $\zeta$ Puppis, our focus is exploring long term changes in the wind and its X-ray heating

\begin{figure}
    \centering
    \includegraphics[width=\linewidth]{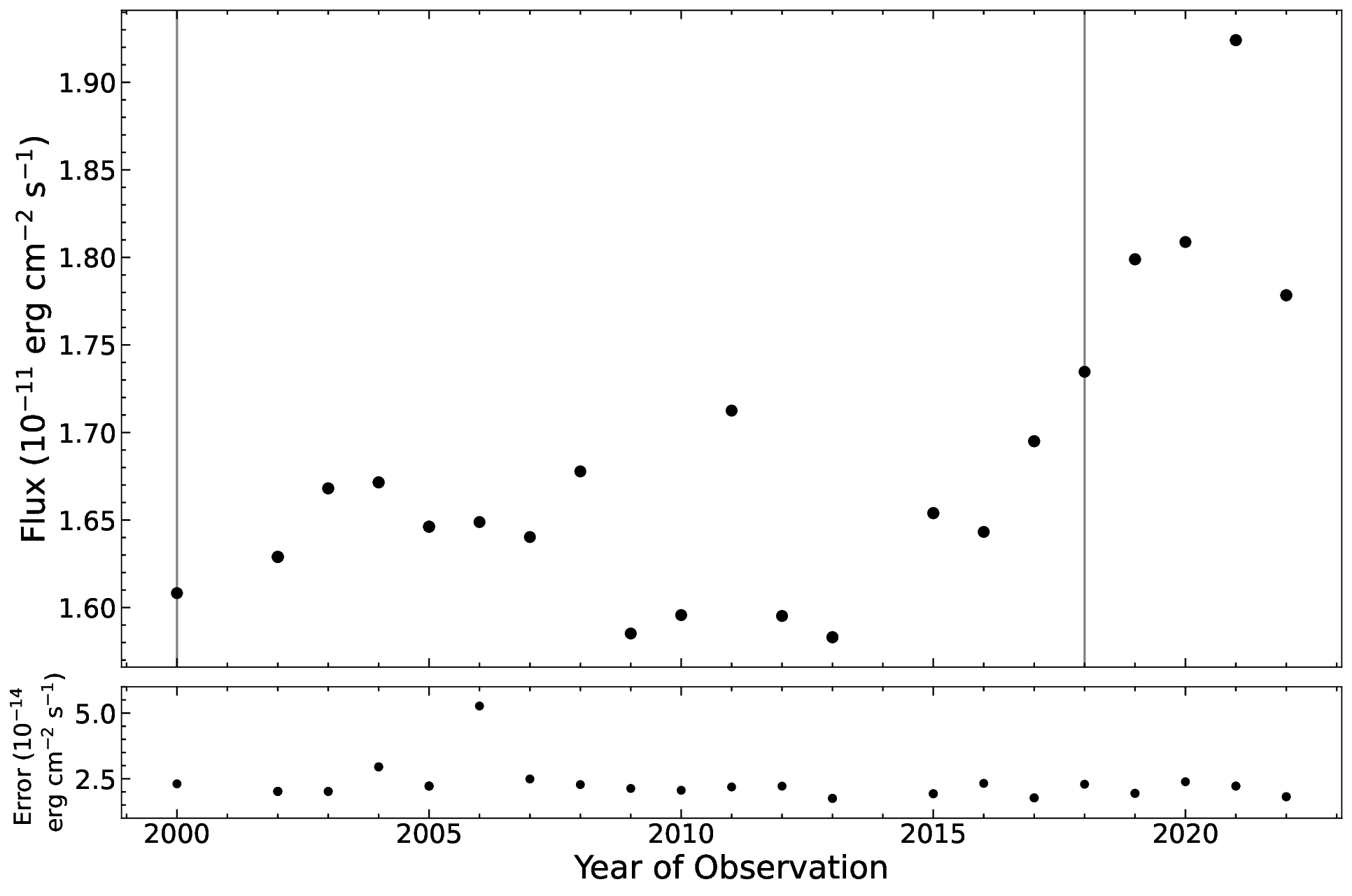}
    \caption{Year-to-year X-ray flux of $\zeta$ Pup measured by \textit{XMM-Newton's} EPIC detector. Vertical lines denote the Cycle 1 \textit{Chandra} observation and the start of the Cycle 19 campaign (see Table~\ref{tab:ObsLog} for dates). Values were retrieved from 4XMM-DR13 \citep{Webb20}. The reported errors in the catalogue are too small to visualize as vertical bars so are plotted in the sub-plot.}
    \label{fig:zPNewtonBrightening}
\end{figure}

The first piece of evidence for changes in $\zeta$ Pup as an X-ray source is its brightening in the last few years as detected with both \textit{Chandra} \citep{Cohen20} and \textit{XMM-Newton}. The latter is shown in Figure~\ref{fig:zPNewtonBrightening}, where the plotted flux values were measured by \textit{XMM-Newton}'s EPIC detector and reported in 4XMM-DR13 \citep{Webb20} \citet{Cohen20} proposed that the increase in X-ray flux is due to the wind mass-flux increasing (more on this immediately below), but the exact origin is not entirely clear due to the the many factors which influence overall X-ray output from embedded wind shocks. More mass within the wind would mean more clumps and subsequently more shocks, but the increase in flux could also be caused by stronger shocks producing more light or a wind that is more porous to X-rays due to changes in clump size or density.

The second is changes in the wind as inferred from a line shape analysis, something that can only be measured using the higher spectral resolution of \textit{Chandra} (though the time coverage is much less frequent). Constraints on the physical effects listed above can be extracted from the X-ray spectral line shapes via some appropriate parameterization of its features. Understanding what the increased flux and changes to the line shapes allude to requires an appropriate line model to parameterize quantities of interest. The inference of the physical changes to the wind using a line profile parameterization is a goal of this paper, for which the chosen parameterization will be discussed in \S~\ref{sec:VBCModel}.

An important parameterization comes from \citet{Owocki01}, which \citet{Cohen10,Cohen20} used to calculate the mass-loss rate of $\zeta$ Pup from X-ray lines measured by \textit{Chandra}. In this paper we will be referring to this model as simply the ``Cohen model" to account for the number of papers that have made use of it \citep{Owocki01,Cohen10,Cohen14,Cohen20}. The most recent of these papers, \citet{Cohen20} came to the remarkable conclusion that the mass-loss rate of $\zeta$ Pup has increased by 40 per cent in a roughly 20-year time frame, based on the changes to the line shapes (i.e., centroid and width) in \textit{Chandra} High Resolution Transmission Grating System (HETGS) data from Cycles 1 and 19 (see Table~\ref{tab:ObsLog} for observation dates). Such a large change in the mass-loss rate of a star brings to light the need to independently verify that observed profile changes are robustly interpretable as changes in the global wind, rather than in just its hot component, given that the hot component is a small fraction of the total wind \citep{Cohen97,Owocki13,Gayley16} and hence might be more susceptible to larger variations.

\subsection{The Challenge of Inferring Mass-loss Rates}

The problem of determining mass-loss rates of massive stars is not a simple one. In Table~\ref{tab:MasslossRates}, we list a number of calculated mass-loss rates for $\zeta$ Pup in the literature. The theoretical value expected from \citet{Vink2000} is unsurprisingly high as their theory does not account for the clumps that make up the wind. \citet{Puls06} made detailed computations of the clumping-included mass-loss rates using H$\alpha$, IR, mm, and radio, finding values that bracket the predicted value from \citet{Vink2000}. The wide difference between the values is a reflection of differing assumptions of where He recombination starts in the wind. These four wave bands provide precise values for the mass-loss rate, but the specific values determined are still heavily dependent on the assumed clumping factor of the wind.

The X-ray values from \citet{Cohen10} and \citet{Cohen20} are free of this assumption since small scale clumping does not affect the reabsorption of X-rays by the cool wind. This allows measurements of X-ray line profiles to directly parameterize the mass-loss rate (see Section~\ref{sec:VBCModel}). Additionally, if the X-ray lines are fit individually in isolated regions, the (relatively flat) local continuum requires no correction for interstellar absorption, so there is no dependence on the distance to $\zeta$ Pup. Thus there is no need for distance corrections like those made by \citet{Howarth19} to the flux-dependent calculations done by \citet{Puls06}. However, the use of X-rays for inferring mass-loss rates is not without its own problems. \citet[hereafter \citetalias{Gunderson22}]{Gunderson22} determined that these estimates are dependent on the assumptions going into the wind profile model used to fit the data.

\begin{table}
    \caption{Archival values of $\zeta$ Pup's mass-loss rate}
    \label{tab:MasslossRates}
    \begin{tabular}{cc}
        \hline
        Reference & $\dot{M}$ ($10^{-6}$~M$_\odot$~yr$^{-1}$) \\
        \hline
        \citet{Vink2000}$^\mathrm{a}$ & $6.4$\\
        \citet{Puls06}$^\mathrm{b}$ & 4.2, 8.5\\
        \citet{Cohen14}$^\mathrm{c}$ & $1.76^{+0.13}_{-0.12}$ \\
        \citet{Howarth19}$^\mathrm{d}$ & 2.6 \\
        \citet{Cohen20}$^\mathrm{e}$ & $2.47^{+0.09}_{-0.09}$ \\
        \hline
    \end{tabular}\\
    \begin{minipage}{7cm}
    \footnotesize
        \textit{Notes} -- $^\mathrm{a}$ Theoretical predicted value.\\
        $^\mathrm{b}$ Derived from simultaneous H$\alpha$, IR, and radio modelling. The two values correspond to different assumed radii of He recombination. \\
        $^\mathrm{c}$ Derived from X-ray line profile modelling Cycle 1 \textit{Chandra} data.\\
        $^\mathrm{d}$ The published mass loss rate have been scaled to reflect the author's newly constrained distance measurement, \citet[$\Dot{M}\propto d^{3/2}$]{Puls06}.\\
        $^\mathrm{e}$ Derived from X-ray line profile modelling Cycle 19 \textit{Chandra} data.
    \end{minipage}
\end{table}

\subsection{Expected Stability of the Global Wind Mass Flux}

In classical theories of massive stars \citep{CAK,Owocki04}, the mass-loss rate is regarded as a stable quantity due to the nature of the line-driving of the wind. Therefore, assuming one's calculated value can be trusted, a large change in the mass-loss rate may be indcative of an incomplete understanding of the stability of these stars over longer periods. Variations on long timescales is not unheard of for stars; for example the Sun undergoes a 22 year cycle \citep{Russell16}. The disks of Be stars also exhibit similar long timescales \citep{Kee20}.

To know if \citet{Cohen20}'s reported changes are part of a long-period variations, we need much more data to see more than one period to determine if it is indeed even periodic, or stochastic change over a longer time scale. As a first step, however, we need to know if the secular changes that have been highlighted in our above discussion correspond to real changes in the global wind properties.

Thus is the goal of this paper. We will investigate the X-ray line shapes of $\zeta$ Pup for the Cycle 1 and 19 \textit{Chandra} observations to investigate the large change in mass-loss rate reported by \citet{Cohen20}. To do so, we will use the VBC model from \citetalias{Gunderson22}. While the specific values of mass-loss rates derived through X-ray lines are dependent on the model used, any large change in a global parameter such as the mass-loss rate should be relativity model independent. To explore this, we will be comparing many of our modelling results with those of \citet{Cohen10} and \citet{Cohen20}, so we will hereafter refer to these works as \citetalias{Cohen10} and \citetalias{Cohen20} respectively.

This paper is organized into the following sections. In \S~\ref{sec:VBCModel}, we give a short summary of the VBC model while including some new insights. In \S~\ref{sec:DataModelling}, we provide details on the data reduction and modelling. Finally, in \S~\ref{sec:Results} and \ref{sec:Conclusions}, we give our results and conclusions.

\section{Variable Boundary Condition Model}\label{sec:VBCModel}

For a more detailed derivation of the VBC model, readers are referred to \citetalias{Gunderson22}. Here we give a summary of the model and its parameters, which starts with the wind's velocity. As with most massive-star wind models, the velocity field is assumed to follow the usual $\beta$-velocity law
\begin{equation}
    v(r) = v_\infty\left(1-\frac{R_*}{r}\right)^\beta.
\end{equation}
We choose $\beta=1$ to allow for analytic results in subsequent equations.  For example, with $\beta = 1$ the optical depth takes the form
\begin{equation}
    \tau(\mu,r;\tau_*)=\tau_*\frac{R_*}{z_t}
    \begin{cases}
    t_-(\mu,r) & \mu\geq0\\
    t_+(|\mu|,r) + \pi & \mu<0
    \end{cases},\label{eq:opticaldepthfull}
\end{equation}
where $\mu=\cos\theta$ and
\begin{align}
    t_\pm &= \arctan{\left(\frac{R_*}{z_t}\right)} \pm \arctan{\left(\frac{\gamma}{\mu}\right)},\label{eq:tau-t}\\
    z_t &= \sqrt{\left(1-\mu^2\right)r^2-R_*^2}\label{eq:zt}\\
    \gamma &= \frac{R_*-r(1-\mu^2)}{z_t}.\label{eq:gammadef}
\end{align}
The variable 
\begin{equation}
    \tau_* \equiv \frac{\kappa(\lambda)}{4\pi R_*v_\infty}\dot{M}\label{eq:taustar}
\end{equation} is a fiducial optical depth describing the amount of absorption a photon produced at $r=R_*$ experiences in a constant velocity wind ($\beta=0$) with the same mass-loss rate as an accelerating wind. For the case of $\beta=1$, it is instead the optical depth at a radius of $r/R_* = \mathrm{e}/(\mathrm{e}-1)\approx1.58$ for a photon on the same central ray.\footnote{This radius interestingly corresponds to the average value of $R_0$ found by \citetalias{Cohen10} and \citetalias{Cohen20}} For information on how $\tau_*$ can be used to calculate a mass-loss rate, see \citetalias{Cohen10}. Additionally, details on the role of complex numbers in the derivations of Equations~\eqref{eq:opticaldepthfull}--\eqref{eq:gammadef} are given in Appendix~\ref{sec:OptDepthReal}.

The main distinguishing feature of the VBC model is its direct parameterization of the heating of the gas. This is in contrast to the use of a hot-gas filling factor, a number which implicitly conflates the effects of the heating and cooling rates. This is achieved through tracking pockets of gas that are assumed to follow a slightly faster velocity field with terminal velocities of $v_{\infty,f}$ than the ambient, slower pockets. Note that this velocity is not used in the definition of $\tau_*$ because it is the ambient, unshocked gas described by $v_\infty$ that absorbs the photons generated in shocks.
The faster clumps are assumed to have a differential probability
\begin{equation}
    \dv{P}{r}=\frac{1}{\ell_0}\label{eq:dPdr}
\end{equation}
of shocking with a slower clump. The above equation contains the final model parameter $\ell_0$ which parameterizes the mean free path between shocks. Thus for a given line can be described by how much optical depth $\tau_*$ the photons traveled through, a characteristic formation radius $R_*+\ell_0$, and the terminal speed $v_{\infty,f}$ of the faster clump that caused the shocked.

The line shape described by these parameters is modelled using the relative luminosity per frequency bin
\begin{equation}
    L(\xi)=\int_{r_{m}(\xi)}^\infty \dv{\mu}{\xi}\mathrm{e}^{-(r-R_*)/\ell_0}\mathrm{e}^{-\tau(\xi,r;\tau_*)}\dv{P}{r}\dd r, \label{eq:luminositybase}
\end{equation}
where $\xi \equiv -\mu v(r)/v_{\infty,f}$ is the frequency shift from line center in terminal speed units and
\begin{equation}
    \dv{\mu}{\xi} = \frac{1}{2}\frac{v_{\infty,f}}{v(r)}.
\end{equation}
is the local mapping from solid angle to frequency space. The lower bound of the integral is the minimum radius capable of producing enough Doppler shift to reach the $\xi$ in question. It is a complicated function dependent on whether the emission is toward the forward ($\mu>0,\xi < 0$) or backward ($\mu<0,\xi>0$) hemisphere of the star. Readers are encouraged to see \citetalias{Gunderson22} for details on this minimum radius. In Appendix~\ref{sec:GL_Approx}, we give details on approximating Equation~\eqref{eq:luminositybase} with a Gauss-Laguerre quadrature.

\subsection{Terminal Velocity Parameter}
In \citetalias{Gunderson22}, a constant, frozen fast-gas terminal velocity of $v_{\infty,f}=2500$ km s$^{-1}$ for each line was assumed. This was done to simplify the model fitting. We use the same process in this work but provide a more mathematical backing for this choice.

Let us assume that the fast gas terminal velocity is a simple scaling constant with respect to the slow gas, $v_{\infty,f}=a v_\infty$. Then, for a given pocket of fast gas that shocks within a mean-free path, its velocity is
\begin{equation}
    v_f = v_{\infty,f} \left(1-\frac{R_*}{R_*+\ell_0}\right) = a v_\infty \frac{\ell_0}{R_*+\ell_0} \sim v_\infty \frac{a\ell_0}{R_*}
\end{equation}
Thus the true parameter to investigate is $a\ell_0$. This is difficult to interpret and compare with existing work however, as other line profile models only parameterize the slow gas terminal velocity. It is for this reason that we choose to fix the fast-gas terminal velocity, i.e. assume a fixed $a$, and vary $\ell_0$ to better compare against spatial parameters in other models.

\subsection{Source Function Description}

\begin{figure}
    \centering
    \includegraphics[width=\linewidth]{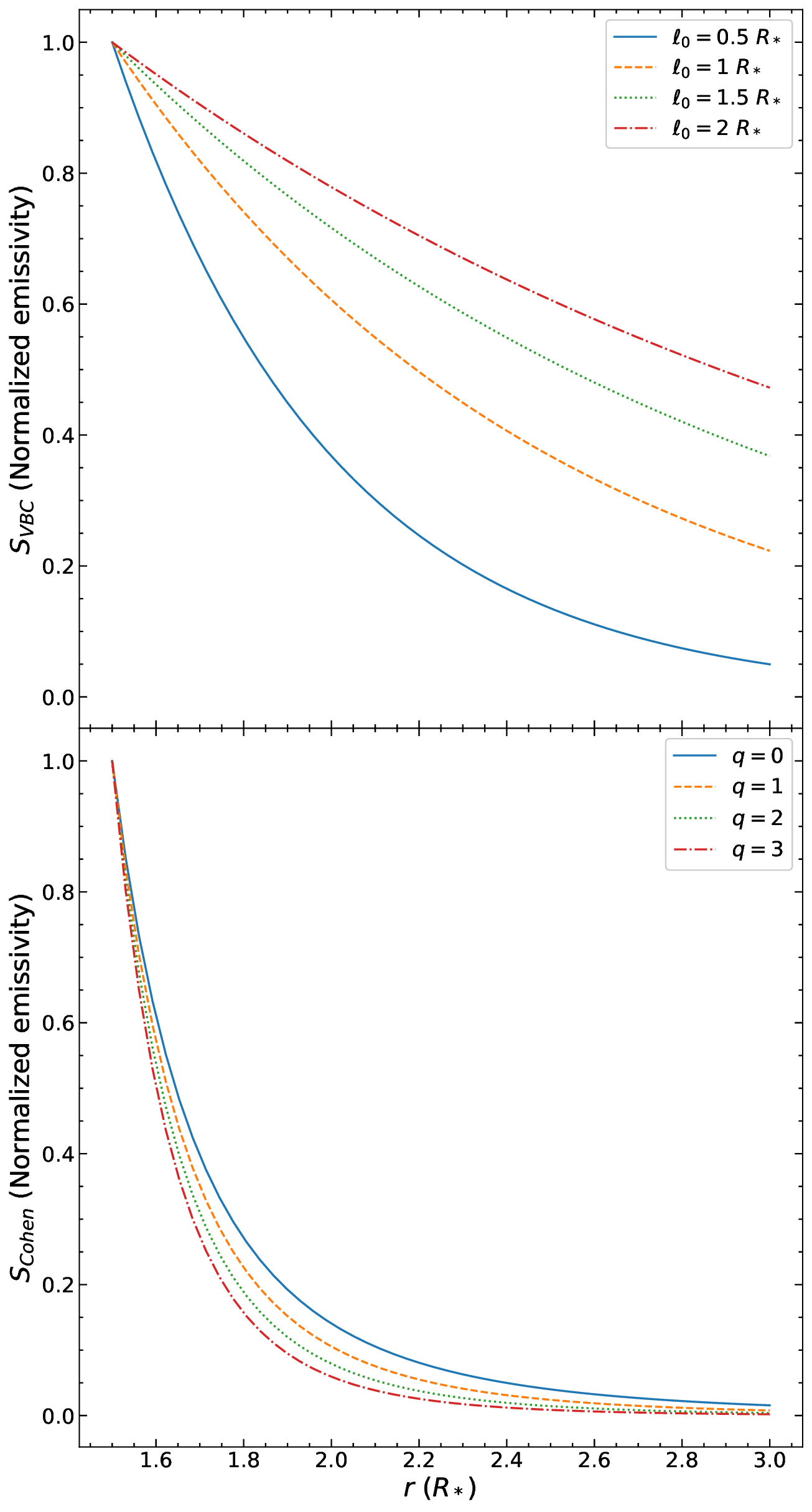}
    \caption{Plot of the line emissivities for the VBC model (top panel) and Cohen model (bottom panel; \citealp{Owocki01}), both starting at a radius of $1.5 R_*$. Note that for the VBC model, the equivalent of the emissivity function Equation~\eqref{eq:VBCemiss} is describing emission over a distance $\ell_0$, not a local emissivity. Both $S_\mathrm{VBC}$ and $S_\mathrm{Cohen}$ are normalized at $r=1.5 R_*$.}
    \label{fig:source}
\end{figure}

In this section we will discuss how the VBC model compares with previously published models when put into the standard language. In general, a line profile function should have the form
\begin{equation}
    L(\lambda) = \int_{V_\mathrm{min}}^\infty S(\lambda, V)\mathrm{e}^{-\tau(\lambda,V)}\dd V,
\end{equation}
where $S(\lambda,V)$ is the emissivity function of the light produced in the volume $V$ of the wind. An example of this $S$ function is the emissivity $\eta_\lambda$ used in the Cohen model. If we change the integration of Equation~\eqref{eq:luminositybase} to that of volume, it follows that our model's equivalent emissivity function (when converted back to radial form) is
\begin{equation}
    S_\mathrm{VBC} = \dv{P}{r}\mathrm{e}^{-(r-R_*)/\ell_0}.\label{eq:VBCemiss}
\end{equation}
Example curves for this emissivity function are plotted for various $\ell_0$ values in Figure~\ref{fig:source}.

For comparison, in the bottom half of that figure we have plotted the emissivity function of the Cohen model. The full emissivity function is given in \citet{Owocki01}, but for our purposes, for a photon emitted at line center, it scales as
\begin{equation}
    S_{\mathrm{Cohen}} = \eta_\lambda \propto \rho^2(r)f(r),
\end{equation}
where $f$ is an X-ray emission filling factor. The authors of this model assume, based on the work of \citet{Ignace01}, that this filling factor is a power law $f(r)\propto r^{-q}$. Calculations of $S_\mathrm{Cohen}$ as a function of radius are shown in the bottom plot of Figure~\ref{fig:source} for a range of values of $q$. The velocity function $v(r)$ which appears in $\rho = \dot{M} / 4 \pi v (r) r^2 $ is calculated using a $\beta$-law with the value $\beta = 1.$

Figure~\ref{fig:source} highlights an important difference between these model parameterizations: the extent of the X-ray emission. Using a constant or slowly varying filling factor causes the emission to stem from a rather narrow radius interval where the cool wind emission measure is high. But since the VBC approach parameterizes the shock heating and assumes efficient radiative cooling, the cool wind emission measure is of no consequence, and the X-ray generation can be more extended if the shock distribution is more extended. Ultimately, only observations and more detailed simulations can clarify which approach better characterizes the spatial distribution of the heating. Therefore our current intent is to understand how robust the conclusions inferred from line profile shapes are when working within these two schemes.

\section{Data and Modelling}\label{sec:DataModelling}

The observations of $\zeta$ Pup included all archival HETGS data available in the \textit{Chandra} Archive. The list of observations with their Observation ID (Obs ID) are given in Table~\ref{tab:ObsLog}. Each observation used the \textit{Chandra} HETGS instrument which simultaneously provides data from two grating arrays: the medium energy grating (MEG) and high energy grating (HEG). These gratings have resolutions of 0.023~\AA\ and 0.012~\AA\ respectively \citep{HETGS_citation}.

\begin{table}
\caption{\textit{Chandra} HETGS Observations of $\zeta$ Pup.}\label{tab:ObsLog}
\begin{tabular}{lcl}
    \hline
    Obs ID & Exposure Time (ks) & Date\\
    \hline
    640$^\mathrm{a}$ & 67.74 & 2000 Mar 28\\
    21113$^\mathrm{b}$ & 17.72 & 2018 Jul 1\\
    21112$^\mathrm{b}$ & 29.70 & 2018 Jul 2\\
    20156$^\mathrm{b}$ & 15.51 & 2018 Jul 3\\
    21114$^\mathrm{b}$ & 19.69 & 2018 Jul 5\\
    21111$^\mathrm{b}$ & 26.86 & 2018 Jul 6\\
    21115$^\mathrm{b}$ & 18.09 & 2018 Jul 7\\
    21116$^\mathrm{b}$ & 43.39 & 2018 Jul 8\\
    20158$^\mathrm{b}$ & 18.41 & 2018 Jul 30\\
    21661$^\mathrm{b}$ & 96.88 & 2018 Aug 3\\
    20157$^\mathrm{b}$ & 76.43 & 2018 Aug 8\\
    21659$^\mathrm{b}$ & 86.35 & 2018 Aug 22\\
    21673$^\mathrm{b}$ & 14.95 & 2018 Aug 24\\
    20154$^\mathrm{b}$ & 46.97 & 2019 Jan 25\\
    22049$^\mathrm{b}$ & 27.69 & 2019 Feb 1\\
    20155$^\mathrm{b}$ & 19.69 & 2019 Jul 15\\
    22278$^\mathrm{b}$ & 30.51 & 2019 Jul 16\\
    22279$^\mathrm{b}$ & 26.05 & 2019 Jul 17\\
    22280$^\mathrm{b}$ & 25.53 & 2019 Jul 20\\
    22281$^\mathrm{b}$ & 41.74 & 2019 Jul 21\\
    22076$^\mathrm{b}$ & 75.12 & 2019 Aug 1\\
    21898$^\mathrm{b}$ & 55.70 & 2019 Aug 17\\
    \hline
\end{tabular}
\begin{minipage}{5.75cm}
	\footnotesize
    \textit{Notes} -- $^\text{a}$ CXO Proposal Cycle 1 observation.\\
    $^\text{b}$ CXO Proposal Cycle 19 observation.
	\end{minipage}
\end{table}

Each observation was retrieved from the \textit{Chandra} archive and reprocessed using the standard pipeline in \textsc{ciao} version 4.13 \citep{Fruscione_ciao}. This process produced two first-order spectra for both the HEG and MEG arrays, corresponding to the positive and negative diffraction orders. Each of the positive and negative orders were subsequently co-added to produce a single first order data set for their respective grating array and Obs ID. For the Cycle 19 datasets, we additionally summed the 42 datasets (21 for the HEG and MEG respectively) to produce a single total observation for each of individual grating array. Thus in the end we had 1 HEG and 1 MEG spectra for Cycle 1 and 1 HEG and 1 MEG spectra for Cycle 19. These final datasets were then rebinned by a constant factor of 3 for model fitting. This corresponds to bins sizes of 0.015~\AA\ for the MEG and 0.0075~\AA\ for the HEG.

When comparing data taken by Chandra separated by almost two decades, we needed to account for changes in the detector properties during that time. The most important change is a result of contamination build up on \textit{Chandra's} ACIS-S detector used for HETGS observations \citep{Herman04,ODell17}. This contamination decreases the detector's effective area, particular in the long-wavelength region of HETGS spectra, reducing overall count rates. However, this change in the detector should not have an important affect on the analysis carried out here. The contamination is calibrated in the standard data products (i.e., date-dependent response files), and we have useful diagnostics out to 17~\AA\ as indicated in Figure~\ref{fig:taustarplot}. In addition, it is worthwhile to note that the Cycle 19 observations have a much greater exposure time, so even with the greater contamination effects there are many more total counts in most of the X-ray lines compared to the Cycle 1 observation.

Using \textsc{xspec} version 12.12 \citep{Arnaud_xspec}, through the \textsc{python} wrap around \textsc{pyxspec}, we fit the same list of lines as in \citetalias{Gunderson22} in isolated regions. The total source model used consisted of the VBC model\footnote{Source code for implementing the model locally can be found here: \url{https://github.com/sjgunderson/VBC_Line_Model}} discussed above plus a constant continuum component. For the line features \ion{Si}{xiii} He $\alpha$,\footnote{In this paper, we shorten the H- and He-like Lyman transition names for conciseness. For example, \ion{Si}{xiii}} He-like Lyman $\alpha$ is shortened to \ion{Si}{xiii} He $\alpha$. \ion{Mg}{xi} He $\alpha$, \ion{Fe}{xvii} at 15.01~\AA, and \ion{Fe}{xvii} at 16.78~\AA, a total of three line model components were used to model the three lines that occur in close spacing at these wavelengths. The $\tau_*$ and $\ell_0$ were the same for each line of these multi-component fits, being tied to the line with the shortest wavelength of the feature's fitted region. The shortest wavelength emission line was chosen for being the strongest single line in these features. Each line component in a feature had a freely fit normalization $N$. We assumed Poisson statistics and used a Cash (maximum likelihood) statistic \citep{Cash79}.

\subsection{Model Fitting Behavior}\label{sec:ModelBehavior}

The fitting algorithm used for our analysis was the Markov chain Monte Carlo (MCMC) function provided in \textsc{pyxspec}. For each fit, the resulting chain was analyzed for all steps after a convergence was achieved, giving parameter distributions like those in Figure~\ref{fig:CornerPlot} in Appendix~\ref{sec:DisExp} for the \ion{Mg}{xii} H $\alpha$ line. This line is within an isolated region with little continuum, so it provides a well constrained example of the model behavior,

There are two interesting trends that can be seen in the parameter pairs. First is the inverse relationship between $\tau_*$ and $\ell_0$. This is similar to the discussion in \citetalias{Gunderson22} when comparing observable-parameter pair contours. A more interesting effect is between $\ell_0$ and the normalization, which show a slight positive correlation. This is likely due to the factor of $1/\ell_0$ that applies to the entire integral in Equation~\eqref{eq:luminositybase} through the relationship defined in Equation~\eqref{eq:dPdr}. Since such a constant factor would normally be considered a part of the normalization, it is not surprising that these would be linked in this way.

\section{Results}\label{sec:Results}

The results of our modelling are summarized in Table~\ref{tab:fitresults} where we report the best-fit values of free parameters in our model. Note that for line features fit with multiple components we report the total flux in the feature. These best-fit values were extracted from the MCMC chains after they stabilized along with their 68 per cent confidence intervals. The actual model fits are plotted in Appendix~\ref{sec:FitPlots}. Figures~\ref{fig:taustarplot} and \ref{fig:l0vsR0} show these parameter results visually.

\begin{table*}
    \caption{VBC model fitting results for Cycles 1 and 19.}
    \begin{tabular}{lcccccccc}
        Line & $\lambda_{\mathrm{p}}$ (\AA)$^\mathrm{a}$ & $\log(T_{\mathrm{max}})$ (log(K))$^\mathrm{b}$ & \multicolumn{2}{c}{$\tau_*$} & \multicolumn{2}{c}{$\ell_0$ $(R_*)$} & \multicolumn{2}{c}{Flux ($10^{-5}$ ph cm$^{-2}$ s$^{-1}$)} \\
        & & & Cycle 1 & Cycle 19 & Cycle 1 & Cycle 19 & Cycle 1 & Cycle 19 \\
        \hline
        \ion{Si}{xiii} He $\beta^\text{c}$ & 5.681 & 7.07 & -- & $0.68^{+0.32}_{-0.20}$ & -- & $1.18^{+0.42}_{-0.16}$ & -- & $0.90^{+0.07}_{-0.07}$ \\
        
        \ion{Si}{xiv} H $\alpha$ & 6.180 & 7.40 & $0.17^{+1.37}_{-0.004}$ & $0.26^{+0.10}_{-0.06}$ & $1.03^{+4.34}_{-0.15}$ & $0.85^{+0.14}_{-0.10}$ & $0.75^{+0.19}_{-0.15}$ & $1.05^{+0.05}_{-0.05}$ \\
        
        \ion{Si}{xiii} He $\alpha^\text{d}$ & 6.648, 6.688, 6.740 & 7.03 & $0.34^{+0.13}_{-0.10}$ & $0.51^{+0.06}_{-0.05}$ & $0.96^{+0.17}_{-0.09}$ & $1.35^{+0.07}_{-0.05}$ & $12.63^{+0.45}_{-0.40}$ & $15.18^{+0.16}_{-0.14}$ \\
        
        \ion{Mg}{xii} H $\alpha$ & 8.419 & 7.19 & $0.75^{+0.35}_{-0.23}$ & $0.60^{+0.07}_{-0.06}$ & $0.82^{+0.29}_{-0.15}$ & $1.18^{+0.08}_{-0.07}$ & $3.00^{+0.31}_{-0.28}$ & $3.84^{+0.10}_{-0.09}$ \\
        
        \ion{Mg}{xi} He $\alpha^\text{d}$ & 9.169, 9.231, 9.314 & 6.84 & $0.75^{+0.18}_{-0.18}$ & $0.60^{+0.07}_{-0.07}$ & $1.15^{+0.20}_{-0.11}$ & $1.69^{+0.12}_{-0.09}$ & $19.67^{+0.70}_{-0.60}$ & $22.96^{+0.28}_{-0.31}$ \\
        
        \ion{Ne}{x} H $\beta$ & 10.238 & 6.97 & $1.63^{+0.76}_{-0.31}$ & $1.84^{+0.24}_{-0.22}$ &  $0.82^{+0.47}_{-0.12}$ & $0.88^{+0.16}_{-0.09}$ & $3.62^{+0.58}_{-0.39}$ & $3.59^{+0.23}_{-0.22}$ \\
        
        \ion{Ne}{ix} He $\beta$ & 11.544 & 6.64 & $0.54^{+0.48}_{-0.18}$ & $1.34^{+0.23}_{-0.16}$ & $2.41^{+1.58}_{-0.43}$ & $1.46^{+0.23}_{-0.04}$ & $6.07^{+0.76}_{-0.82}$ & $7.07^{+0.40}_{-0.37}$ \\
        
        \ion{Ne}{x} H $\alpha$ & 12.132 & 6.94 & $1.41^{+0.19}_{-0.15}$ & $1.71^{+0.12}_{-0.11}$ & $1.03^{+0.15}_{-0.10}$ & $1.32^{+0.11}_{-0.07}$ & $26.92^{+1.36}_{-1.45}$ & $27.24^{+0.80}_{-0.81}$ \\
        
        \ion{Fe}{xvii}$^\text{d}$ & 15.014, 15.176, 15.261 & 6.73 & $1.48^{+0.22}_{-0.18}$ & $1.57^{+0.16}_{-0.15}$ & $1.27^{+0.17}_{-0.10}$ & $1.75^{+0.15}_{-0.10}$ & $79.13^{+3.81}_{-3.01}$ & $92.54^{+2.55}_{-2.62}$ \\
        
        \ion{Fe}{xvii}$^\text{d}$ & 16.780, 17.051, 17.096 & 6.70 & $2.75^{+0.67}_{-0.37}$ & $3.14^{+0.80}_{-0.49}$ & $1.28^{+0.19}_{-0.14}$ & $2.06^{+0.43}_{-0.27}$ & $62.67^{+2.95}_{-2.91}$ & $82.40^{+4.23}_{-3.73}$ \\
        \hline
    \end{tabular}
    \label{tab:fitresults}
    \begin{minipage}{17cm}
	\footnotesize
    \textit{Notes} -- $^\text{a}$Predicted line center wavelength as reported by AtomDB \citep{Foster_AtomDB,Smith_AtomDB}.\\
    $^\text{b}$Temperature of maximum emissivity as reported by AtomDB \citep{Foster_AtomDB,Smith_AtomDB}.\\
    $^\text{c}$This line is not confidently detected in the Cycle 1 data.\\
    $^\text{d}$The reported flux for these line features is the total flux in the three components used.
	\end{minipage}
\end{table*}

\subsection{Determination of Mass-loss Rate}

\begin{figure*}
    \centering
    \includegraphics[width=\linewidth]{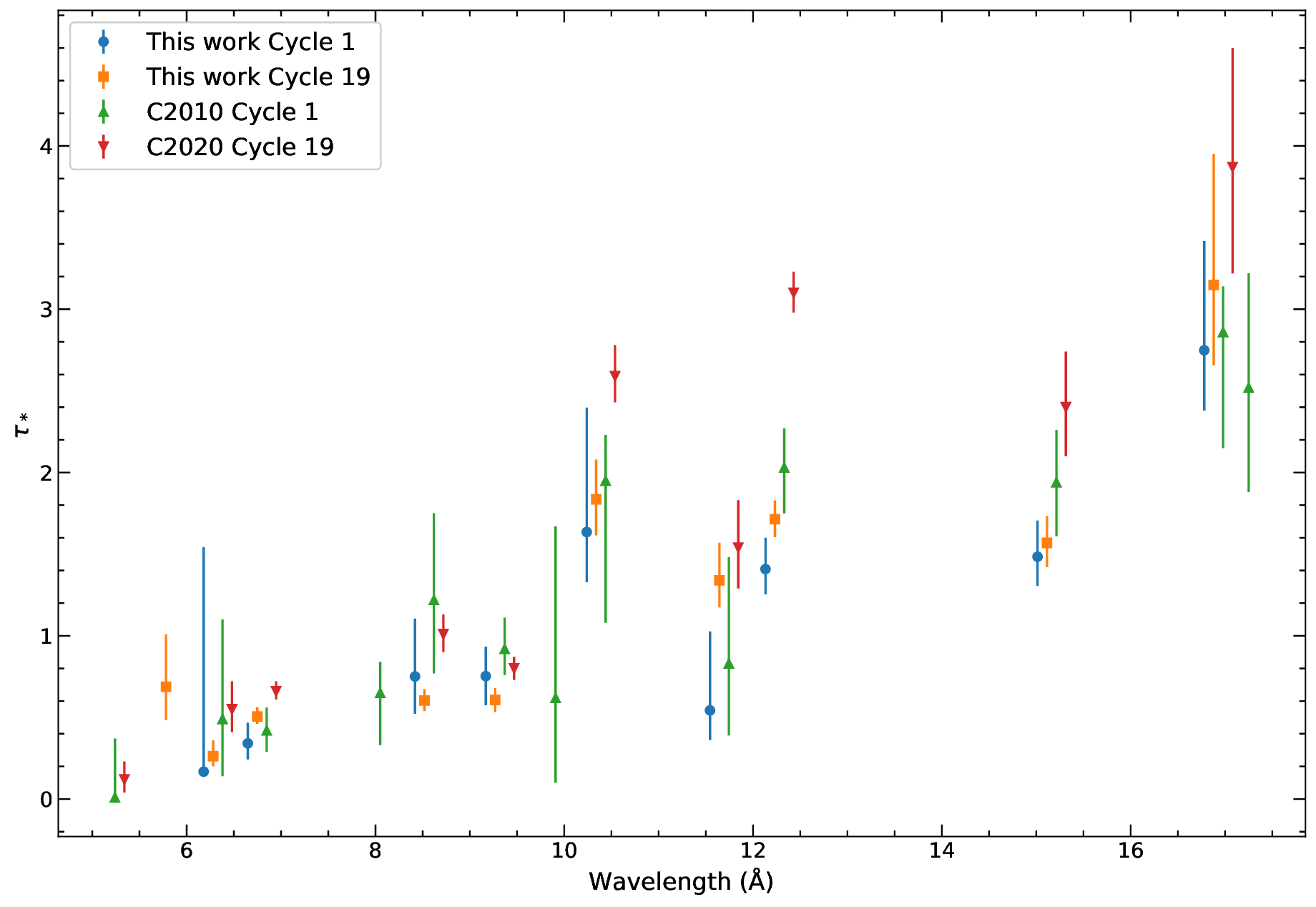}
    \caption{Plot of best-fit $\tau_*$ values for the four analyses discussed. Error bars represent 68 per cent uncertainty. Note that we have applied constant offsets to the different result groups to make the error bars more visible. The plotted wavelengths correspond to $\lambda_\mathrm{p}$ for the blue dots, $\lambda_\mathrm{p}+0.1$ \AA\ for the orange squares, $\lambda_\mathrm{p}+0.2$ \AA\ for the green up triangles, and $\lambda_\mathrm{p}+0.3$ \AA\ for the red down triangles. Data points that do not have a corresponding group of four are due to that line feature not being fit in some of the works considered here.}
    \label{fig:taustarplot}
\end{figure*}

Focusing first on Figure~\ref{fig:taustarplot}, we can draw two important conclusions. First, it appears that mass-loss rates calculated from X-ray line profiles is independent of the assumed heating model. The best-fit values for our Cycle 1 analysis (blue dots) and \citetalias{Cohen10} (green up-triangles) show an over-all consistency when looking at 68 per cent confidence intervals. This is further bolstered by the values from our Cycle 19 analysis (orange squares) as the consistency is maintained. For all three fit groups, we would find the same mass-loss rate, since $\tau_*$ is a simple rescaling of $\Dot{M}$ based on Equation~\eqref{eq:taustar}, save for deviations accounted for by errors in a trend-line fitting process. 

Given this overall consistency, the second conclusion we draw from Figure~\ref{fig:taustarplot} is that $\zeta$ Pup's mass-loss rate has \textit{not} changed. If such a change occurred, particularly one suggested to be so large, our independent analysis on the same datasets would have found a similar result. However, we do confirm that there is a real change in the lines. The flux in all but one of our fitted line features has increased, just as was found by \citetalias{Cohen20}. Combined with the \textit{XMM-Newton} flux in Figure~\ref{fig:zPNewtonBrightening}, there is high confidence in more X-rays being produced and/or escaping the wind.

\subsection{Issues with Epochal Analysis}

The result that mass-loss rate determinations from X-ray lines is model independent is in contrast to \citetalias{Gunderson22} that concluded the opposite by analyzing only Cycle 19 data. The choice in only looking at Cycle 19 data may have been the source of the erroneous conclusion as \citetalias{Gunderson22} was comparing against the results of \citetalias{Cohen20}, which is the only set to break from the noted consistency. Why this is the case is not immediately clear. For lines below 10~\AA, all four groups are in agreement except for \ion{Si}{xiii} He $\alpha$, though only marginally so. For the longer wavelengths, there are large differences in the values found for the Cohen model. It is possible that the limited fitting region of the \ion{Ne}{x} H $\alpha$ and \ion{Fe}{xvii} at 15~\AA\ lines used by \citetalias{Cohen20} is responsible for some of the inflated values of $\tau_*$. The fits presented here contrast this by fitting the entire red-half of the profile of \ion{Ne}{x} H $\alpha$ and all three lines in the \ion{Fe}{xvii} at 15~\AA\ feature (Figures~\ref{fig:Cycle1Fits3} and \ref{fig:Cycle19Fits3}). This is simply speculation as to what differences may exist between the two modelling approaches that would result in the different $\tau_*$ values found. Whatever the source is, there may be anomalous behavior in the \citetalias{Cohen20} fitting analysis that is also responsible for their claim of a 40 per cent increase in $\zeta$ Pup's mass-loss rate.

\begin{figure}
    \centering
    \includegraphics[width=\linewidth]{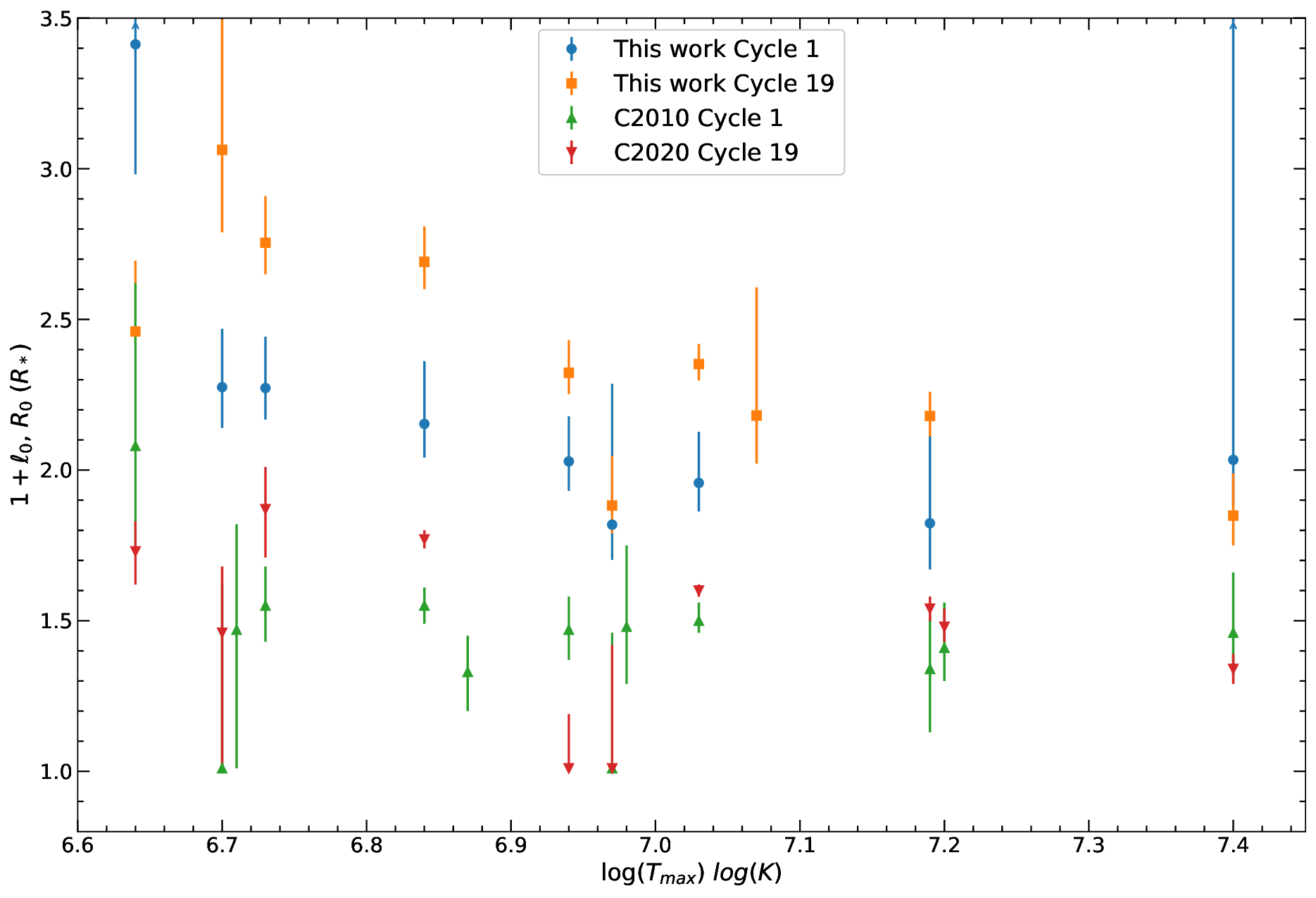}
    \caption{Plot of best-fit spatial parameters values for the four analyses discussed against the temperatures of maximum emissivity taken from the APED. The VBC model's parameter is plotted as $R_* + \ell_0$ in units of stellar radii. Error bars represent 68 per cent uncertainty. Points with large errors are not plotted in entirety for readability and are represented by error bars with arrows.}
    \label{fig:l0vsR0}
\end{figure}

From their values in Table~\ref{tab:fitresults}, we argue that this difference stems from the $\ell_0$ parameter, which is also the source of the line shape differences that were noted by \citetalias{Cohen20}. Put another way: the distribution of the embedded wind shocks has changed. Figure~\ref{fig:l0vsR0} more clearly shows that the mean-free path between shocks in $\zeta$ Pup's wind has increased by a significant amount. Note that in this figure we plot $R_* + \ell_0$ so as to compare directly with the turn-on radius $R_0$ of the Cohen model. What this means is that the shocks are occurring farther out in the wind, allowing for more X-ray emission to escape, thereby raising the amount of flux that we see. 

This presents two competing interpretations of the same data: either the mass-loss rate has changed or the shock distribution. In comparing these, it is worth considering the implications each has over the entire wind. Given \citetalias{Cohen20}'s increased mass-loss rate, the entire wind is required to be changed to account for the differences we see in the line profile shapes. Without evidence of $\zeta$ Pup having large scale variabilities, i.e. those that persist after time-averaging during an observation, that alter the stellar parameters, such a large increase in the mass-loss rate would imply either that $\zeta$ Pup has undiscovered properties or our understanding of massive stars, particularly O-types, is missing an important piece. Given the importance of $\zeta$ Pup as a prototype, guide star, and calibration target, it has many further implications in every other wave band. The conclusion from the VBC model analysis that only the shock distribution has changed, and not the global mass-loss rate, has less drastic implications involving a much smaller fraction of the wind.

\subsection{Shock-Temperature Spatial Distribution}

The primary motivation for using the VBC analysis is to understand how the gas is heated, not to determine the cool wind mass loss rate. So now we turn to the issue of using the VBC parameterization to interpret the shock distribution, and contrast that with the picture from Cohen's parametrization. For example, part of the motivation of using a turn-on radius $R_0$ in the Cohen model is the idea that the line deshadowing instability requires a certain stand-off distance in order to have room to take effect, whereas the VBC picture holds that fast gas is destined to overtake slow gas over the length scale $\ell_0$. In the latter picture, it is natural that more significant boundary variations should yield more violent encounters between fast and slow wind, all at lower $\ell_0$, whereas in the former picture, there could be more of a tendency for $R_0$ to be independent of shock strength, or even rise with shock strength \citep{Driessen21}.

\citetalias{Gunderson22} visually identified a trend in $\ell_0$ vs. the temperature of maximum emissivity $T_\mathrm{max}$. That trend can be demonstrated quantitatively using the Pearson correlation coefficients
\begin{equation}
    \rho_{xy} = \frac{n \sum x_iy_i - \sum x_i\sum y_i}{\sqrt{n\sum x_i^2 - \left(\sum x_i\right)^2}\sqrt{n\sum y_i^2 - \left(\sum y_i\right)^2}}
\end{equation}
for the plotted points in Figure~\ref{fig:l0vsR0}. The values of the coefficients $\rho_{\ell_0T_\mathrm{max}}$ are given in Table~\ref{tab:TrendTable}. Note that this correlation coefficient is invariant under linear transformations, so the calculated values are the same for both $\ell_0$ and $R_* + \ell_0$. Additionally, we calculated $\rho_{\ell_0T_\mathrm{max}}$ using $\log(T_\mathrm{max})$, but for readability of the coefficient subscripts, we omit the logarithm.

We used a random sampling method of the $\ell_0$ parameter distributions to calculate the reported values. The central values are the most frequent of the resulting $\rho_{\ell_0T_\mathrm{max}}$ distribution while the uncertainties are then taken as the 68 per cent levels.
 
We are interested in investigating trends related to the shock strength, for which the temperature of maximum emissivity $T_\mathrm{max}$ is a useful proxy. From the values in Table~\ref{tab:TrendTable}, a negative correlation does exist between the $\ell_0$ and the shock strength for both Cycle 1 and Cycle 19. The trend is not exactly linear, which would correspond to $\rho_{\ell_0T_\mathrm{max}}=-1$, but it does imply an inverse trend nonetheless.

Of note is how the Cycle 1 results were handled. Two of the lines fit with our model, \ion{Si}{xiv} H $\alpha$ and \ion{Ne}{ix} He $\beta$, show significantly more uncertainty in their best fit $\ell_0$ compared to the rest of the lines. When these points are included, the correlation is consistent with 0, meaning no linear relationship exists. However, the rest of the Cycle 1 results show a tighter trend with $\log(T_\mathrm{max})$, so we also calculated the correlation coefficients with these two lines removed. The restricted coefficient is constrained to only negative values and is statistically consistent with the computed value with all lines considered. If the uncertainties in the outlier lines can be better constrained, the true value is likely to be within the overlap of these two values. These coefficients confirm the prediction that under a VBC initiation for a massive star's wind, the stronger shocks will occur at lower radii.

\begin{table}
    \caption{Pearson correlation coefficients.}
    \begin{tabular}{lc}
        Dataset & $\rho_{\ell_0T_\mathrm{max}}$ \\
        \hline
        \multirow{2}{*}{Cycle 1} & $-0.56_{-0.06}^{+1.02}$$\;^\mathrm{a}$ \\
        & $-0.82_{-0.06}^{+0.42}$$\;^\mathrm{b}$ \\
        Cycle 19 & $-0.75_{-0.07}^{+0.11}$ \\
        \hline
    \end{tabular}
    \label{tab:TrendTable}
    \begin{minipage}{5cm}
	\footnotesize
    \textit{Notes} -- $^\text{a}$Correlation coefficient when outlier values from \ion{Si}{xiv} H $\alpha$ and \ion{Ne}{ix} He $\beta$ are included.\\
    $^\text{b}$Correlation coefficient without outlier values from \ion{Si}{xiv} H $\alpha$ and \ion{Ne}{ix} He $\beta$.
	\end{minipage}
\end{table}

There are interesting ramifications for how the shock heating distribution appears to have changed with time. While $R_0$ parameter in the Cohen model is broadly unchanged, the $\ell_0$ parameter of the VBC model has increased significantly. Such a difference between the temporal changes in the shock distributions is likely explained from the fact that in the VBC model it is only the hot gas that has changed. This requires a more significant adjustment of the heating distribution instead of the entire wind. The latter of which is required by the Cohen model based on the wind mass flux increase \citepalias{Cohen20}. Thus we can conclude that the hot gas has moved to farther radii in recent years while the cool, unshocked gas has not changed, though some changes can not be ruled out.

\section{Conclusions}\label{sec:Conclusions}

We present new and better constrained results for the VBC line model fit to both epochs of X-ray line emission data on $\zeta$ Pup as observed by \textit{Chandra}. Based on the best-fitting values of the $\tau_*$ parameter, we find that $\zeta$ Pup's mass-loss rate has not undergone any large-scale changes within the near two decade between of the two observations. Additionally, the consistency between our Cycle 1 and 19 $\tau_*$ values and those of \citetalias{Cohen10} implies that mass-loss rates inferred using X-ray line emission are independent of model parameterizing and thus robust as a technique.

It is concerning then that a 40 per cent change in the mass-loss rate shows up in one parameterization and not in the other. However, rather than reflecting an overall incompatibility between the two models, the difference is possibly from the specifics of the line analysis. For example, there appears to be a dependence on whether an entire line feature, such as the three resolved lines of the \ion{Fe}{xvii} at 15~\AA, is fit versus a single line of the feature. This is based on our modelling accounting for the three lines in this feature for both Cycle 1 and 19 and finding a smaller $\tau_*$ value than \citetalias{Cohen10}  and \citetalias{Cohen20}, the former still being within error, which fit only the \ion{Fe}{xvii} line at 15.014~\AA. It remains unclear what systematic uncertainties are present in the mass flux calculations due to the complexity of the modelling choices and ad-hoc nature of the heating distributions used in the two models discussed. Further work should be done on two fronts to confront these problems. First, the source of uncertainties in modelling should be investigated as the potential source for the discrepancies noted in this work. Secondly, models with more self-consistent shock physics should be developed to better approach this problem.

If indeed $\zeta$ Pup’s mass-loss rate has not changed, we must then explain why there are observed increases in the line fluxes that are not due to changes in \textit{Chandra}’s detectors. Our spatial parameter $\ell_0$ provides such an explanation: the embedded wind shocks are occurring at farther radii now than before. The further out a shock occurs, the more photons can escape since they travel through less material. The increase in $\ell_0$ can be traced back to the surface where the lag between the boundary variations is getting longer, giving the slower gas a longer head start before the fast gas is launched. So if the shocks are seeded by boundary fluctuations, this would suggest the timescale of the fluctuations has gotten longer.  That any such change is possible on a 20-year timescale is already interesting, and a challenge to interior models, but some sort of stellar cycle seems potentially relevant. Future monitoring of $\zeta$ Pup in the X-ray band is needed to test the apparent long-term variations.

\section*{Acknowledgements}

The scientific results in this article are based on data retrieved from the \textit{Chandra} Data Archive, software provided by the Chandra X-ray Center (CXC) in the application packages \textsc{ciao}, and software provided by NASA's High Energy Astrophysics Science Archive Research Center (HEASARC) in the application \textsc{xspec}. This research was supported in part through high performance computational resources provided by The University of Iowa, Iowa City, Iowa. This research has made use of data obtained from the 4XMM \textit{XMM-Newton} Serendipitous Source Catalog compiled by the 10 institutes of the \textit{XMM-Newton} Survey Science Centre selected by ESA.

Support for SJG was provided by the University of Iowa's CLAS Dissertation Writing Fellowship and a Chandra X-ray Center (CXC) Research Visitor Award. Support for DPH and SJG was provided by NASA through the Smithsonian Astrophysical Observatory (SAO) contract SV3-73016 to MIT for Support of the CXC and Science Instruments. CXC is operated by SAO for and on behalf of NASA under contract NAS8-03060.

This work was original published in SJG's doctoral thesis in partial fulfillment of the University of Iowa's Graduate College Ph.D. requirements. \citep{GundersonThesis}

We also thank R. Ignace and C. Erba for their discussions on putting the VBC model into context of the usual modelling approaches and interpreting our results.
\section*{Data Availability}

The X-ray spectral data used in this article are available in the \textit{Chandra} Data Archive at \url{https://cxc.cfa.harvard.edu/cda/}. Observations are uniquely identified by an observation identifier (Obs ID) given in Table~\ref{tab:ObsLog}.



\bibliographystyle{mnras}
\bibliography{BIB} 


\appendix

\section{Optical Depth Expression: Real vs. Complex}\label{sec:OptDepthReal}

As was noted in the discussion of the minimum radius for the integral of Equation~\eqref{eq:luminositybase} in \citetalias{Gunderson22}, complex numbers tend to show up due to the roots that are being taken in the formulas in the VBC model. One more source of complex numbers that has not been mention is that of the optical depth in Equation~\eqref{eq:opticaldepthfull}.

To be clear, as a variable corresponding to a physical quantity, $\tau(\mu,r;\tau_*)$ is always real. The problem of complex numbers arises in the sub-equations that make it up. Specifically, $z_t$ from Equation~\eqref{eq:zt} becomes imaginary when $r < R_*/\sqrt{1-\mu^2}$. If one wishes or needs to avoid this in their own implementation of the VBC model equations, the following expanded form of Equation~\eqref{eq:tau-t} can be used
\begin{equation}
    t_\pm = \begin{cases}
    \arctan{\left(\frac{R_*}{z_t}\right)} \pm \arctan{\left(\frac{\gamma}{\mu}\right)} & r\geq\frac{R_*}{\sqrt{1-\mu^2}}\\
    -\left(\mathrm{arctanh}{\left(\frac{R_*}{z_t}\right)} \pm \mathrm{arctanh}{\left(\frac{\gamma}{\mu}\right)}\right) & r<\frac{R_*}{\sqrt{1-\mu^2}},
    \end{cases}
\end{equation}
which then requires that
\begin{equation}
    z_t = \begin{cases}
    \sqrt{\left(1-\mu^2\right)r^2-R_*^2} & r\geq\frac{R_*}{\sqrt{1-\mu^2}}\\
    \sqrt{R_*^2-\left(1-\mu^2\right)r^2} & r<\frac{R_*}{\sqrt{1-\mu^2}}.
    \end{cases}
\end{equation}

\section{Gauss-Laguerre Quadrature Approximation}\label{sec:GL_Approx}

The use of a Gaussian quadrature algorithm for calculating numerical integrals is a well-known method of fast and precise approximation, particularly for situations of infinite integration bounds. For the VBC model, the Gauss-Laguerre quadrature is the appropriate option to use when re-written in an applicable form. To do this, we will make the substitution $y = (r-r_m(\xi))/\ell_0$. This turns Equation~\ref{eq:luminositybase} into the form
\begin{equation}
    L(\xi)=\mathrm{e}^{-(r_m(\xi)-R_*)/\ell_0}\int_0^\infty \mathrm{e}^{-y} h(y,\xi)\dd y
\end{equation}
where
\begin{equation}
    h(y,\xi)=\frac{1}{2}\frac{y\ell_0 + r_m(\xi)}{y\ell_0 + r_m(\xi) - R_*}\mathrm{e}^{-\tau(y,\xi)}
\end{equation}
We can then write a semi-analytic expression for the line profile
\begin{equation}
    L(\xi)\approx\mathrm{e}^{-(r_m(\xi)-R_*)/\ell_0}\sum_{i=1}^n w_ih(y_i,\xi).\label{eq:L-GL}
\end{equation}
Here $y_i$ are the $i$-th roots of the classical orthogonal Laguerre polynomial $\mathcal{L}_n$ and the weights are
\begin{equation}
    w_i = \frac{y_i}{(n+1)^2\mathcal{L}_{n+1}(y_i)^2}
\end{equation}

\begin{figure}
    \centering
    \includegraphics[width=\linewidth]{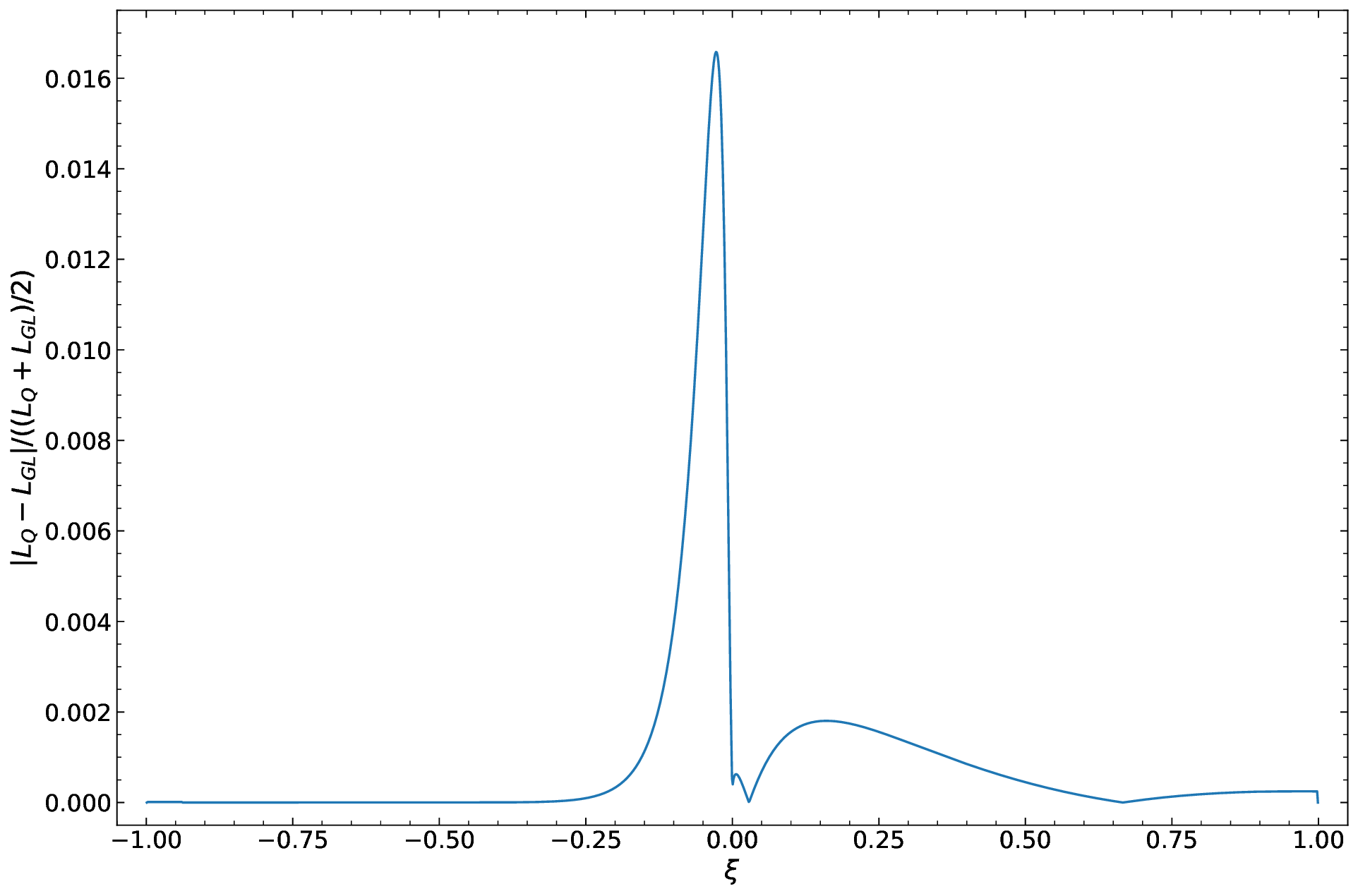}
    \caption{Relative difference between the general purpose quadrature $L_{\mathrm{Q}}$ and Gauss-Laguerre quadrature $L_{\mathrm{GL}}$ integration methods available in the VBC model codes.}
    \label{fig:AlgPerDiff}
\end{figure}

To show that this semi-analytic form does not introduce uncertainties greater than the instrumental sources, we have plotted the relative difference between using \textsc{scipy}'s general purpose quadrature algorithm $L_{\mathrm{Q}}$ and the Gauss-Laguerre algorithm $L_{\mathrm{GL}}$ for $n=20$ in Figure~\ref{fig:AlgPerDiff}. At the greatest, the difference between these two methods are of order 1.7 per cent, far below any uncertainties from statistical sources. Both of these methods are included in the available files for implementing the VBC model on personal distributions of \textsc{pyxspec}.

\section{VBC Parameter Distribution Example}\label{sec:DisExp}

\begin{figure*}
    \centering
    \includegraphics[width=\linewidth]{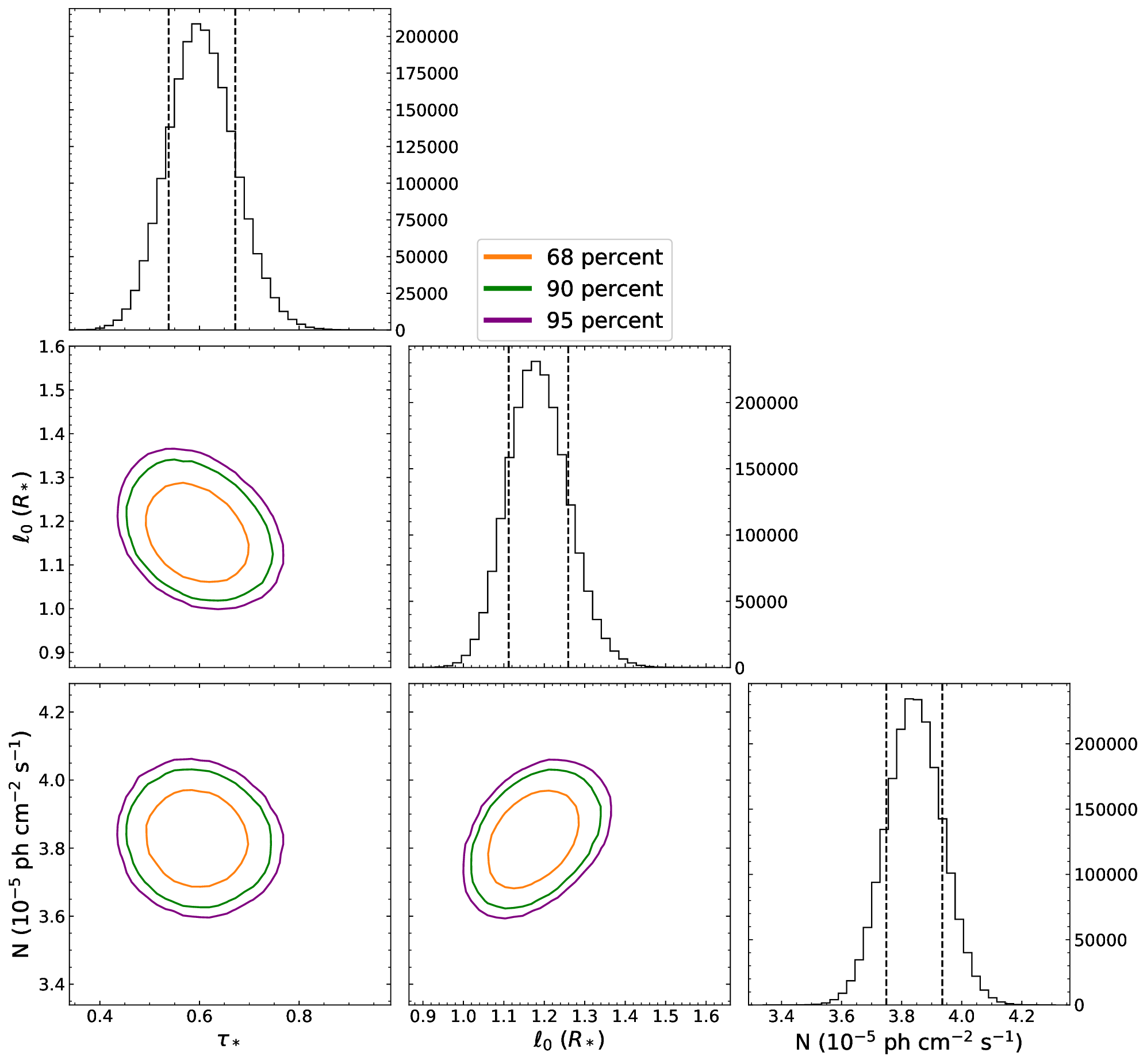}
    \caption{Example parameter-pair confidence contours for the VBC model fit to \ion{Mg}{xii} H $\alpha$ with $v_{\infty,f}$ frozen. Dashed vertical lines in the marginal distributions are the 68 per cent levels for that parameter. Parameter-pair contours are representative of model behavior for most lines fit.}
    \label{fig:CornerPlot}
\end{figure*}

\section{Fit Plots}\label{sec:FitPlots}

\begin{figure*}
    \centering
    \includegraphics[width=\linewidth]{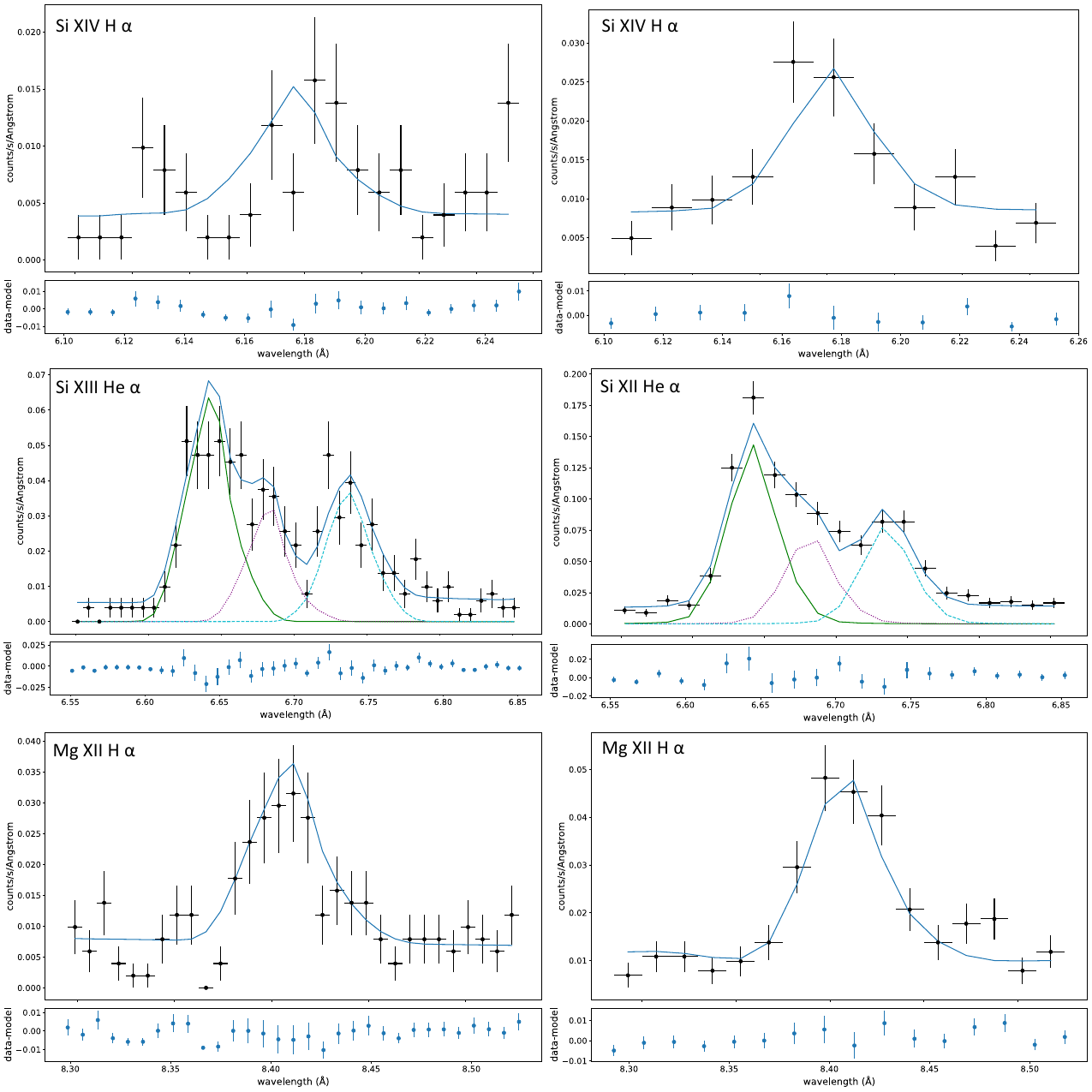}
    \caption{Cycle 1 best fit plots of \ion{Si}{xiv} H $\alpha$, \ion{Si}{xii} He $\beta$, and \ion{Mg}{xi} H $\alpha$. Left column is the HEG data while the right is the MEG. Fits that used more than one component also include the individual components. All plots are shown binned by a factor 3, as was used for the fitting process.}
    \label{fig:Cycle1Fits1}
\end{figure*}

\begin{figure*}
    \centering
    \includegraphics[width=\linewidth]{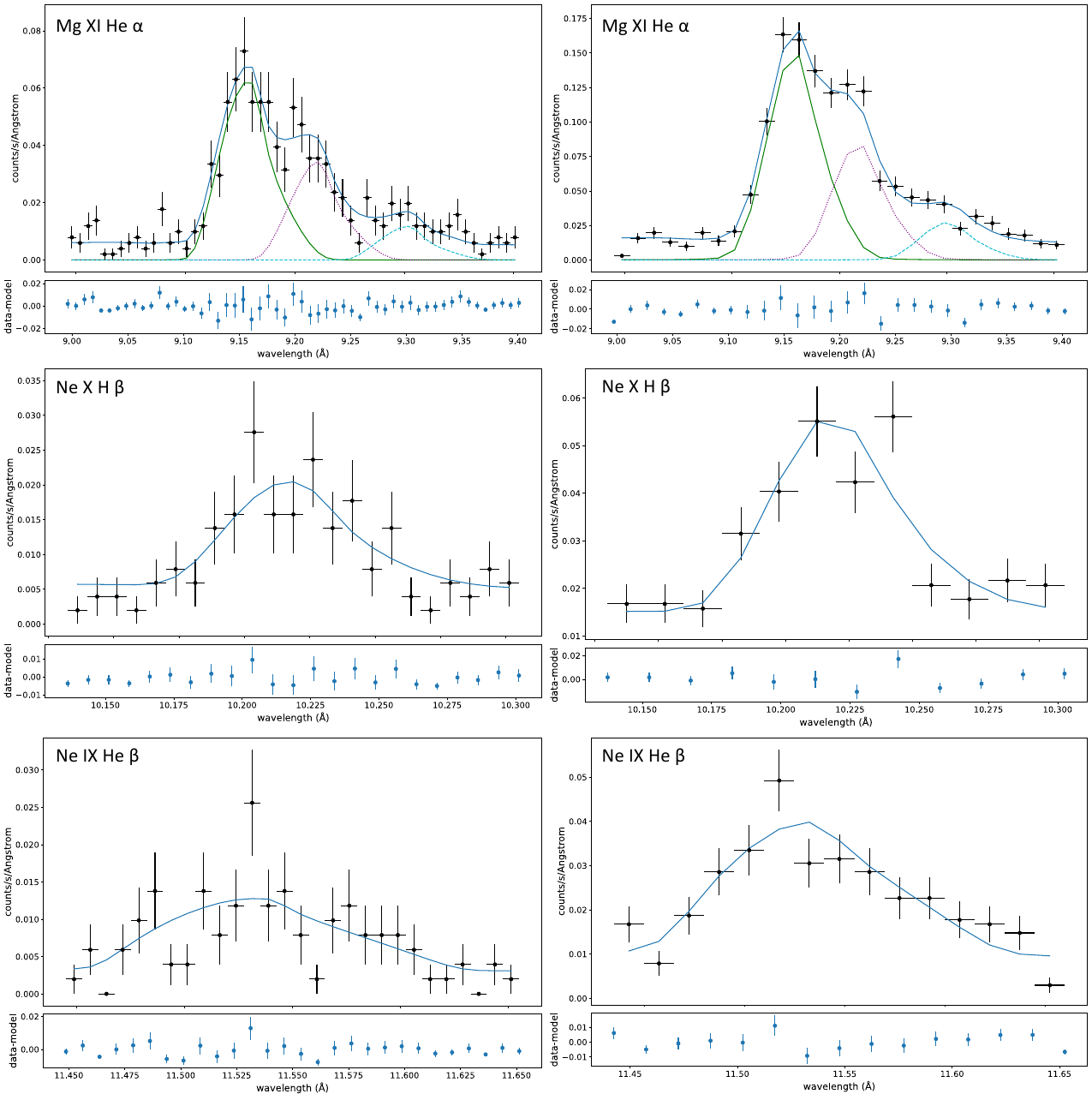}
    \caption{Cycle 1 best fit plots of \ion{Mg}{xi} He $\alpha$, \ion{Ne}{x} H $\beta$, and \ion{Ne}{ix} He $\beta$. Left column is the HEG data while the right is the MEG. Fits that used more than one component also include the individual components. All plots are shown binned by a factor 3, as was used for the fitting process.}
    \label{fig:Cycle1Fits2}
\end{figure*}

\begin{figure*}
    \centering
    \includegraphics[width=\linewidth]{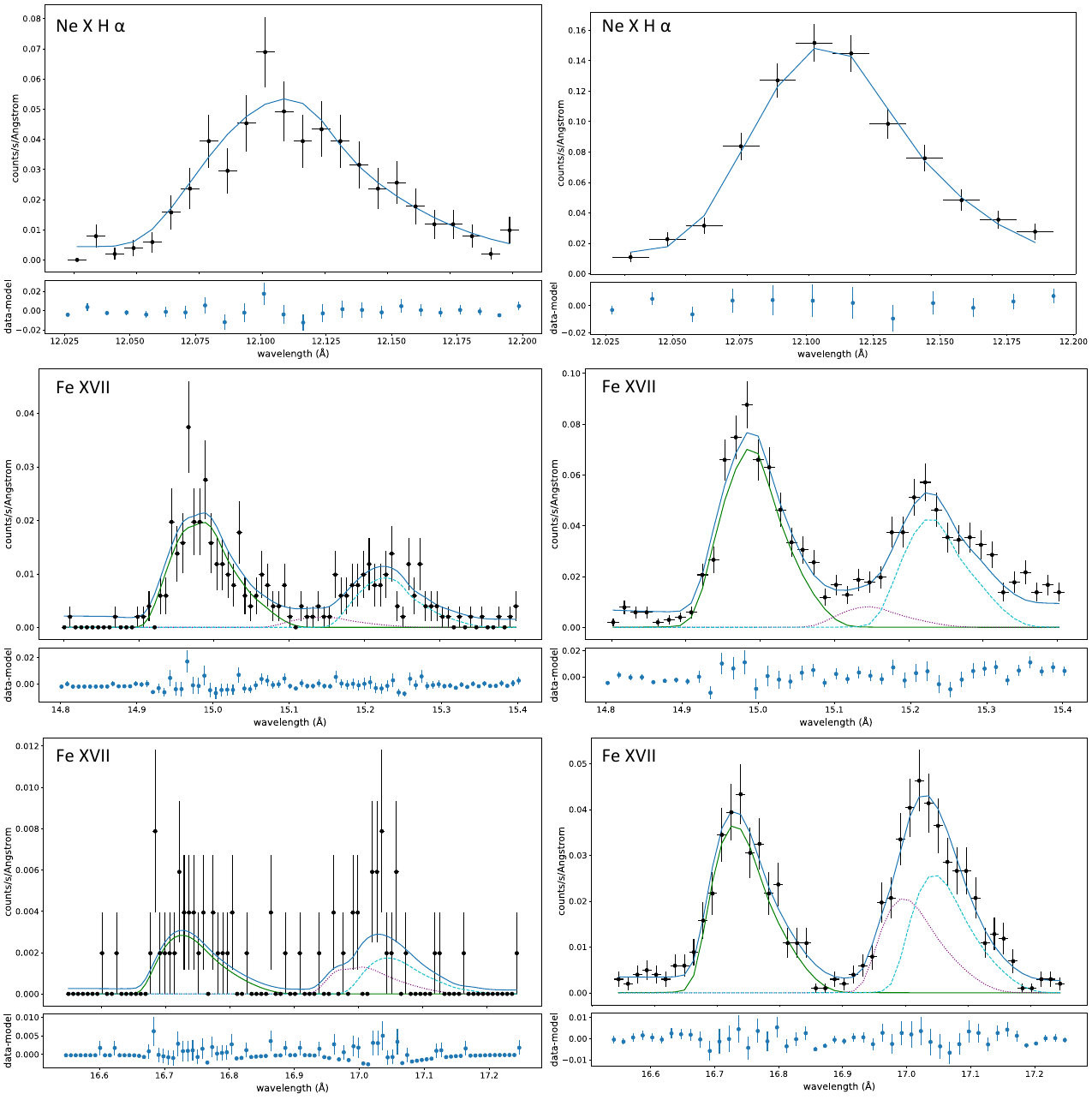}
    \caption{Cycle 1 best fit plots of \ion{Ne}{x} H $\alpha$, \ion{Fe}{xvii} at 15.01 \AA, and \ion{Fe}{xvii} at 16.68 \AA. Left column is the HEG data while the right is the MEG. Fits that used more than one component also include the individual components. All plots are shown binned by a factor 3, as was used for the fitting process.}
    \label{fig:Cycle1Fits3}
\end{figure*}

\begin{figure*}
    \centering
    \includegraphics[width=\linewidth]{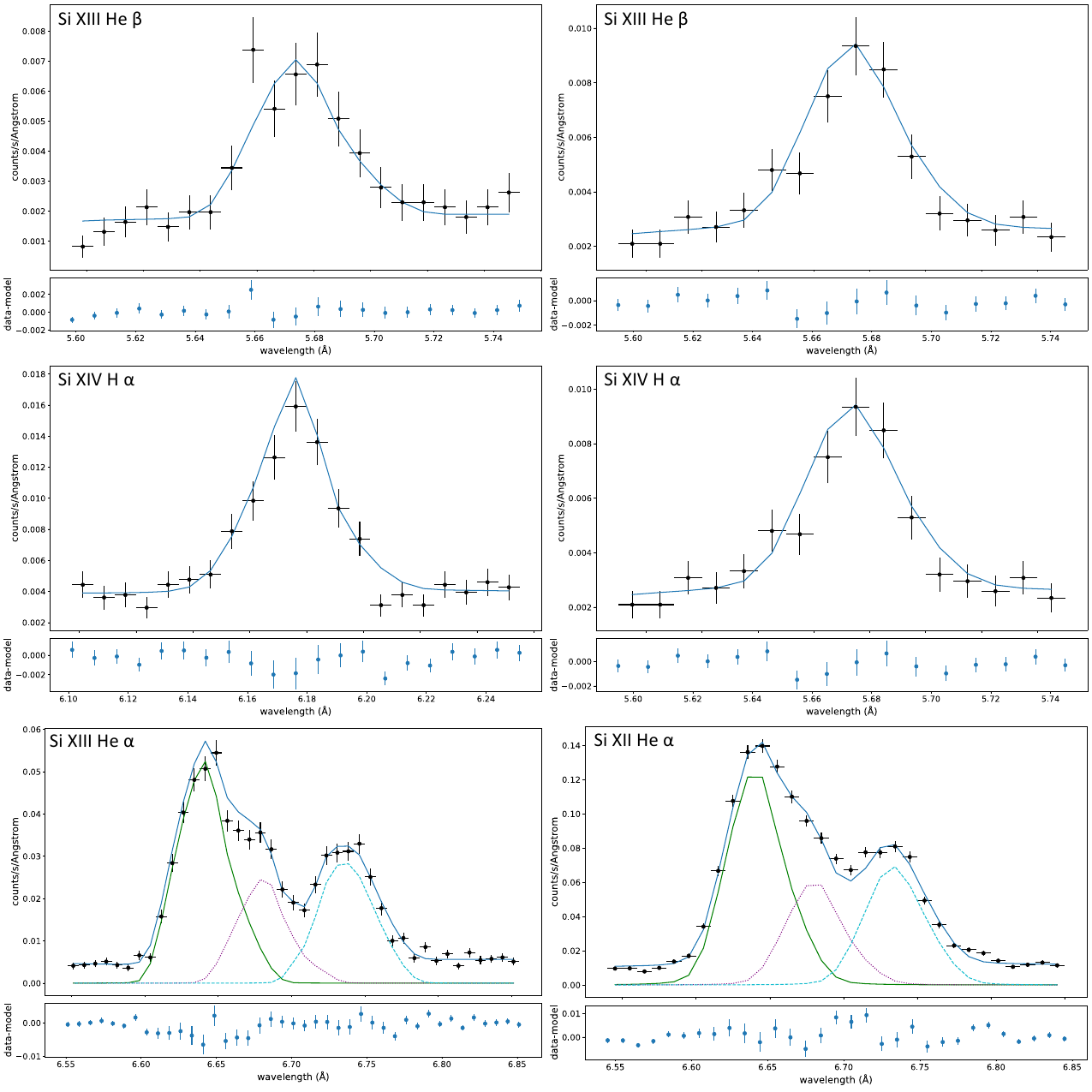}
    \caption{Cycle 19 best fit plots of \ion{Si}{xiii} He $\beta$, \ion{Si}{xiv} H $\alpha$, and \ion{Si}{xiii} He $\alpha$. Left column is the HEG data while the right is the MEG. Fits that used more than one component also include the individual components. All plots are shown binned by a factor 3, as was used for the fitting process.}
    \label{fig:Cycle19Fits1}
\end{figure*}

\begin{figure*}
    \centering
    \includegraphics[width=\linewidth]{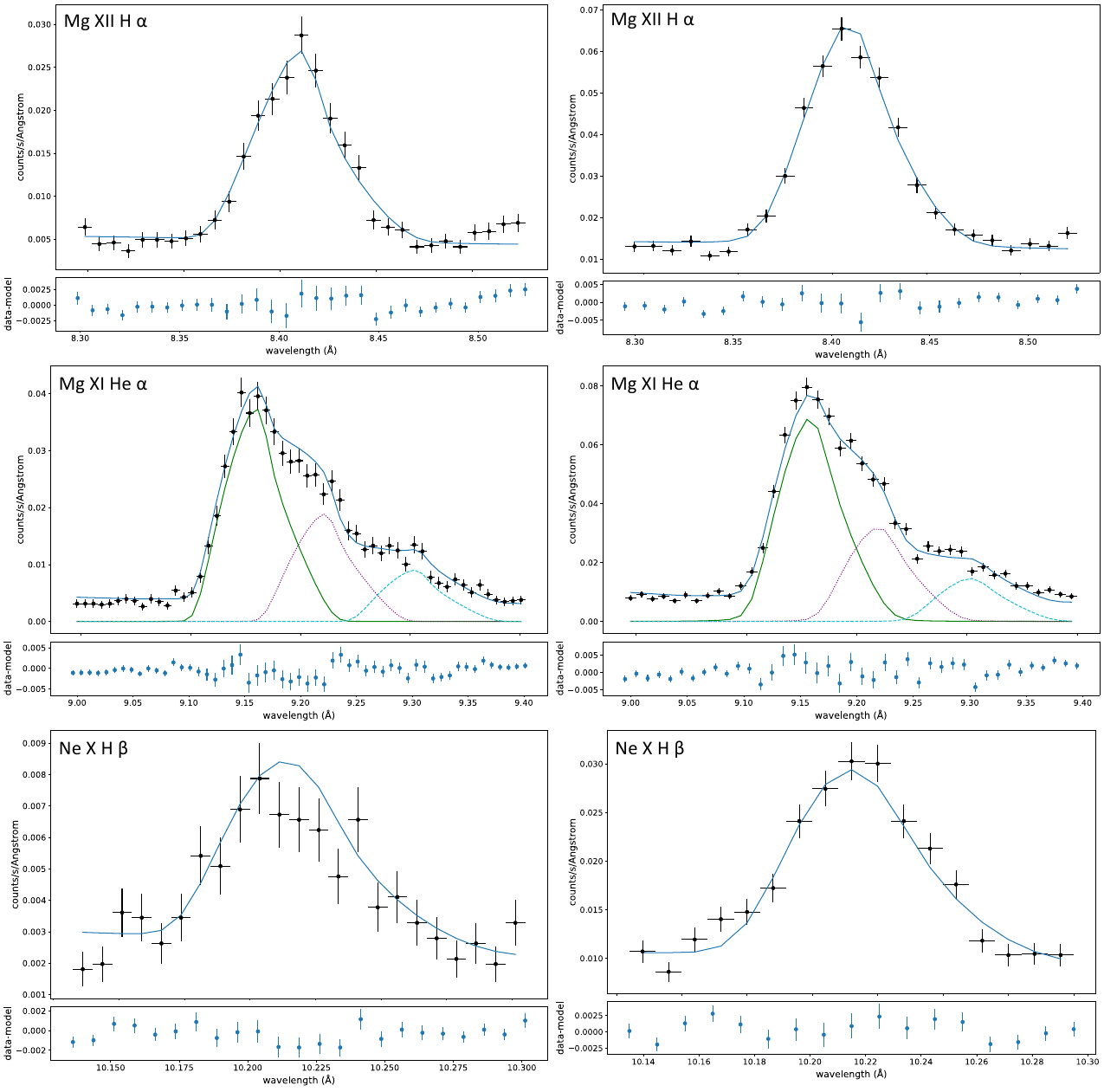}
    \caption{Cycle 19 best fit plots of \ion{Mg}{xii} H $\alpha$, \ion{Mg}{xi} He $\alpha$, and \ion{Ne}{x} H $\beta$. Left column is the HEG data while the right is the MEG. Fits that used more than one component also include the individual components. All plots are shown binned by a factor 3, as was used for the fitting process.}
    \label{fig:Cycle19Fits2}
\end{figure*}

\begin{figure*}
    \centering
    \includegraphics[width=\linewidth]{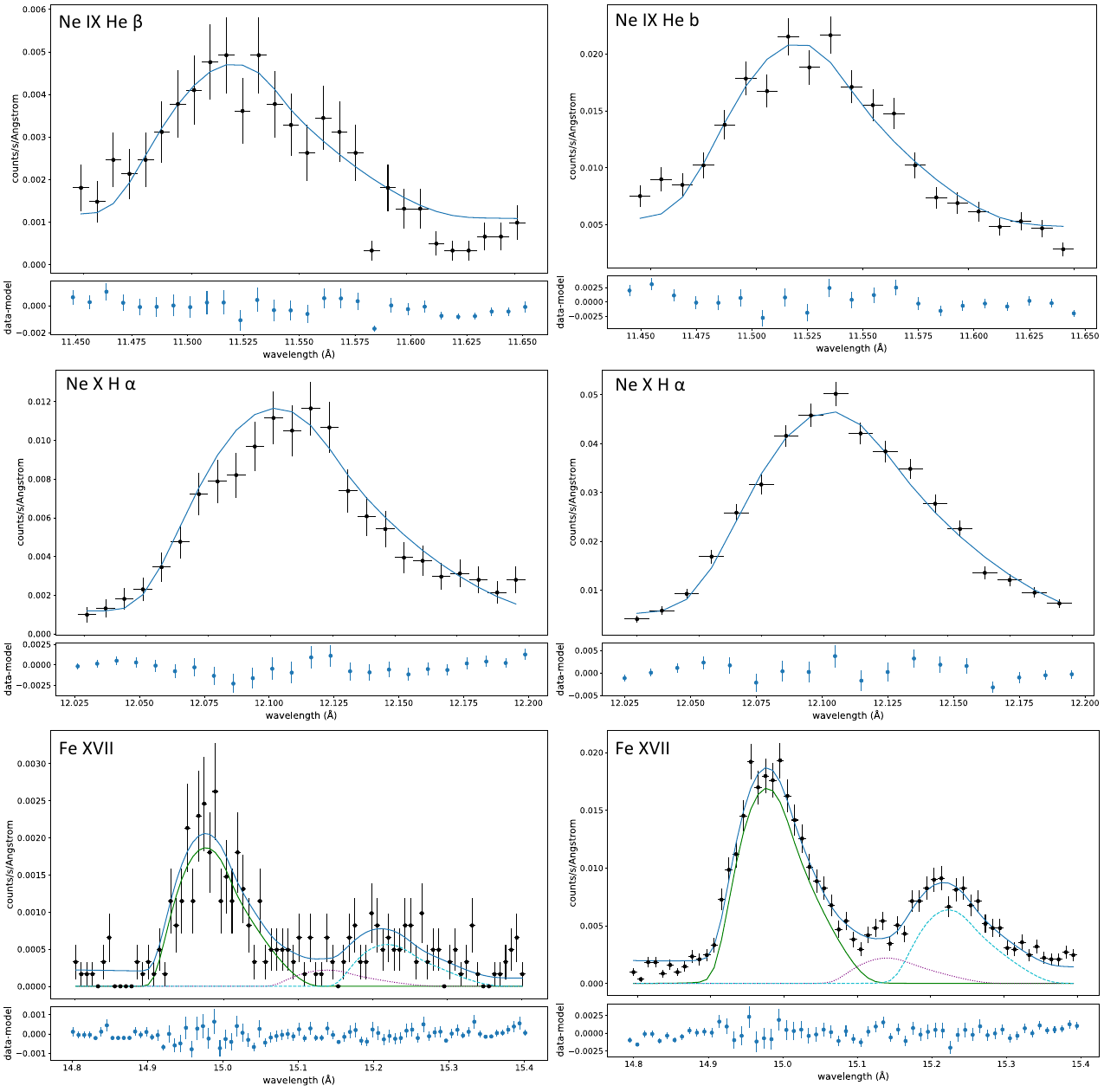}
    \caption{Cycle 19 best fit plots of \ion{Ne}{ix} He $\beta$, \ion{Ne}{x} H $\alpha$, and \ion{Fe}{xvii} at 15.01 \AA. Left column is the HEG data while the right is the MEG. Fits that used more than one component also include the individual components. All plots are shown binned by a factor 3, as was used for the fitting process.}
    \label{fig:Cycle19Fits3}
\end{figure*}

\begin{figure*}
    \centering
    \includegraphics[width=\linewidth]{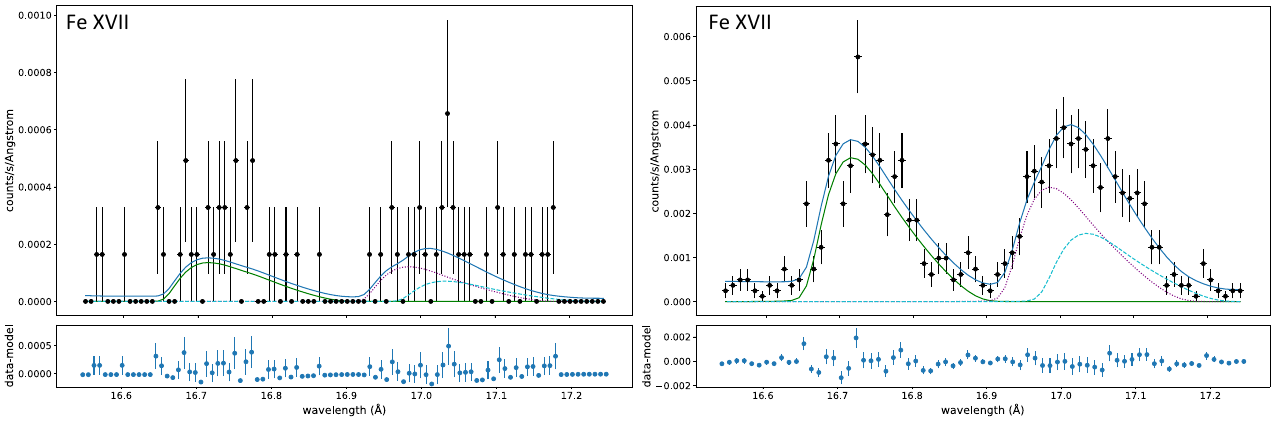}
    \caption{Cycle 19 best fit plots of \ion{Fe}{xvii} at 16.78 \AA. Left column is the HEG data while the right is the MEG. Fits that used more than one component also include the individual components. All plots are shown binned by a factor 3, as was used for the fitting process.}
    \label{fig:Cycle19Fits4}
\end{figure*}


\bsp	
\label{lastpage}
\end{document}